\documentclass{elsart}

\usepackage{graphics}
\usepackage{epsfig}
\usepackage{subfigure}
\usepackage{amsmath,amsfonts,amssymb,graphicx,times}

\begin{document}

\begin{frontmatter}

\title{Ranking the spreading influence in complex networks}

\author[1]{Jian-Guo Liu}\ead{liujg004@ustc.edu.cn},
\author[1]{Zhuo-Ming Ren},
\author[1]{Qiang Guo}

\address[1]{Research Center of Complex Systems Science, University of
Shanghai for Science and Technology, Shanghai 200093, PR China}

\begin{abstract}
Identifying the node spreading influence in
networks is an important task to optimally use the network structure
and ensure the more efficient spreading in information. In this
paper, by taking into account the shortest distance
between a target node and the node set with the highest $k$-core value,
we present an improved method to generate the ranking list to evaluate the
node spreading influence. Comparing with the epidemic process results for four real networks and the Barab\'{a}si-Albert network, the parameterless method could identify the
node spreading influence more accurately than the ones generated by the degree $k$, closeness centrality, $k$-shell and mixed degree
decomposition methods. This work would be helpful
for deeply understanding the node importance of a network.\end{abstract}

\begin{keyword}
Network science \sep Spreading influence \sep K-shell decomposition.
\PACS 89.75.Hc\sep 87.23.Ge\sep 05.70.Ln
\end{keyword}

\end{frontmatter}

\section{Introduction}
\begin{figure}
\center\scalebox{1.3}[1.3]{\rotatebox{180}{\includegraphics{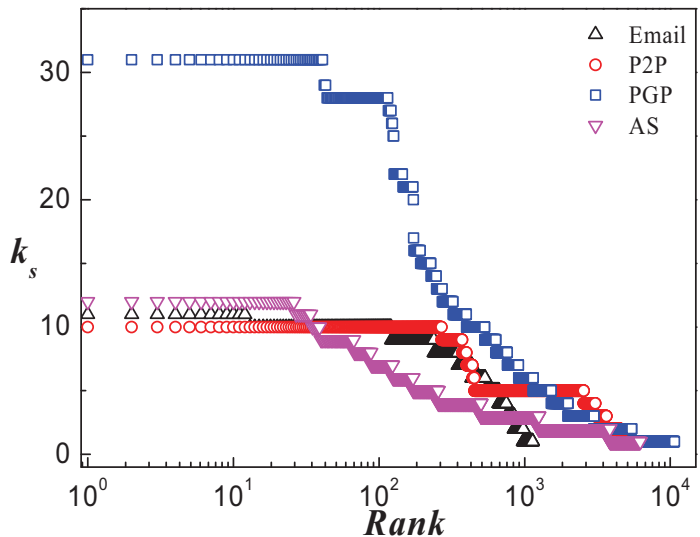}}}
\caption{(Color online) Rank of the $k_s$ values for Email, P2P,
PGP, and AS networks, from which one can find that there are lots of
nodes whose $k_s$ values are equal.} \label{Fig1}
\end{figure}

Spreading is an ubiquitous process in nature, which describes many
important activities in society \cite{1,r01,r02,PRE2006}, such as the virus
spreading \cite{2,3}, reaction diffusion processes \cite{4,jliu2007},
pandemics \cite{5}, cascading failures \cite{6} and so on. The knowledge of the spreading pathways through
the network of interactions is crucial for developing effective
methods to either hinder the disease spreading, or accelerate the
information dissemination spreading. So far, a lot of works focus on
identifying the most influential spreaders in a network \cite{7,r09,r10},
for example, the most connected nodes (hubs) are supposed to be the
key spreaders, being responsible for the largest scale of the
spreading process \cite{8,10,ref10}. Recently,
Kitsak {\it et al.} \cite{1} argued that the node spreading influence is determined by its location in a network. By decomposing a network with the $k$-shell decomposition method, they found that the most
influential nodes, namely the network core, could be identified by the largest $k$-core values. It should be noticed that
the $k$-shell method assigns many nodes with the
the same $k$-core value even though they perform entirely difference in the
spreading process. Figure \ref{Fig1} shows that, for some real networks, there
are lots of nodes whose $k$-core values, denoted as $k_s$, are
equal. By taking into the number of removed and existed links in the decomposition process, Zeng {\it et al.}
\cite{11} proposed an improved method, named mixed degree
decomposition (MDD) method, to distinguish the node spreading
influence within the node set with the same $k_s$ value. To different networks, the optimal parameters of the MDD method
are determined by the statistical properties of the networks, which hinder its application.
By investigating the effects of privileged spreaders on social networks, Borge {\it et al.} \cite{r09,r10}
found that the node spreading influence does not depend on their $k$-core values, which instead determines whether or not a given node
prevents the diffusion process. These literatures suggest that, besides the network core, it is also very important to generate a ranking
list to identify all nodes' spreading influences. In this paper, we argue that,
for the node set with same $k$-core values, the nodes whose locations are close to the network core have larger spreading influences.
Inspired by the idea, we present an improved $k$-shell method to generate the global influential ranking list.
Comparing with the susceptible-infection-recovered (SIR) spreading process \cite{12,13} for four
real networks and the Barab\'{a}si-Albert network \cite{14}, the experimental results show that
our method could generate the ranking list more accurately than the ones generated by the degree $k$, closeness centrality(CC)\cite{cc}, $k$-shell and MDD decomposition methods respectively.

\section{Method}

Normally, a network $G=(N,E)$ with $N$ nodes and $E$ links could be described by an adjacent matrix $A=\{ a_{ij} \}\in R^{n,n}$, where
$a_{ij}=1$ if node $i$ is connected by node $j$, and $a_{ij}=0$ otherwise. The node degree $k_i$ is defined as the number of neighbors for node $i$. The closeness centrality(CC) of node $i$ is defined as the reciprocal of the sum of the shortest distances to all other nodes of N\cite{cc}.The $k$-shell decomposition method \cite{shell01,shell02} could be
implemented in the following way to identify the network core. Firstly, remove all nodes with degree one, and then keep pruning the existed
nodes until all nodes' degrees are larger than one. The removed nodes would form a node set whose $k$-core value
equals to one. Then, repeat the pruning process in the same
way for the rest nodes. Finally, the $k$-shell method decomposes a network into different node set with different $k$-core values.
Implementing the SIR spreading process for one network, one can find that the nodes with
the same $k_s$ values always have different number of infected nodes, namely {\it spreading influence}.
This phenomena suggests that the $k$-shell decomposition method is
not appropriate for ranking the global spreading influence of a network. In terms of the distance from a target node to the network core,
the spreading influences of the nodes with the same $k$-core values could be distinguished in the following way
\begin{eqnarray}\label{eqn1}
\theta(i|k_s)=(k_s^{\rm max}-k_{s}+1)\sum_{j\in J} d_{ij} , \ i\in S_{k_s}.
\end{eqnarray}
where $k_s^{\rm max}$ is the largest $k$-core value of a network,
the shortest distance $d_{ij}$ is measured by the shortest
distance from the node $i$ to the node $j$, $J$ is denoted as the network core node set, and $S_{k_s}$ is denoted as the node set whose $k$-core values equal to $k_s$.

\section{How to evaluate the performance}
To check the performance of the improved method, four real networks are introduced in this paper, which include (\romannumeral1) {\bf Email
network} \cite{16}. the Email network of University Rovira i Virgili
(URV) of Spain contains faculty, researchers,
technicians, managers, administrators, and graduate students.
(\romannumeral2) {\bf Peer-To-Peer (P2P) network} \cite{17}. A
sequence of snapshots of the Gnutella peer-to-peer file sharing
network from August 2002. Each node represents a host in the
Gnutella network and each link represents the connection between each
pair of Gnutella hosts. (\romannumeral3) {\bf Pretty-Good-Privacy
(PGP) network} \cite{18}. Pretty-Good-Privacy algorithm have been
developed in order to maintain privacy between peers, wherefore, it
is also called web of trust of PGP. (\romannumeral4){\bf Autonomous
Systems (AS) network} \cite{19}. The data was collected from
University of Oregon Route Views Project - Online data and reports.
The network of routers comprising the Internet can be organized into
sub-graphs called Autonomous Systems. Each AS exchanges traffic
flows with some neighbors. The statistical properties include the number of nodes $N$ and links $E$ of the network,
the average degree, the second-order average degree, and the spreading threshold are given in Table \uppercase\expandafter{\romannumeral1}.

\begin{table}
\caption{The statistical properties of the networks, where $N$ is the
number of nodes, $E$ is the number of links, $\langle k\rangle=\frac{1}{N}\sum_i k_i$ and $\langle k^2\rangle=\frac{1}{N}\sum_i k_i^2$
are the average degree and the second-order average degree of the network, $\beta^{c}_{\rm rand}$ is
the spreading threshold for a network \cite{epidemic_threshold}
($\beta^{c}_{\rm rand}\approx\langle k\rangle/\langle
k^2\rangle$).}
\begin{center}
\begin{tabular} {lrrrrr}
  \hline \hline
   Networks &$N$    &$E$    &$\langle k\rangle$ &$\langle k^2\rangle$ &~~$\beta^{c}_{\rm rand}$     \\ \hline
   Email    &1133   &5451   &9.60                &180                  &~~0.053     \\
   P2P      &6301   &20778  &7.00                &116                  &~~0.060     \\
   PGP      &10680  &24316  &4.60                &86                   &~~0.053      \\
   AS       &6202   &12170  &3.92                &618                  &~~0.006      \\
   \hline \hline
    \end{tabular}
\end{center}
\end{table}

To evaluate the performance of the improved method, the Kendall's tau
\cite{kendall22,kendall23} is introduced to measure the accuracy of the method.
By using the degree $k$, $k$-shell and MDD methods, we could obtain different ranking lists
in terms of the network structure. In
principle, the ranking lists generated by an effective structure-based ranking method should be
as close as possible to the ranking list generated by the real spreading process.
In this paper, we employ the SIR model \cite{1} to
simulate the spreading process on networks. In the SIR model, we denote that all nodes are initially susceptible
except the only one infectious node $i$. In each time step, the infected
nodes will infect their susceptible neighbors with the spreading rate $\beta$,
and infected nodes would recover in two time steps \cite{1}.
The number of infections $s^\beta_i$ generated by the initially-infected
node $i$ is denoted as its spreading influence, where $\beta$ is the
spreading rate in the SIR model. Ranking the node spreading influence in terms of its spreading influence $s^\beta_i$,
one could obtain the ranking list of the SIR-model-based spreading influence. We
therefore use the Kendall's tau coefficient $\tau$ to measure the correlation between one topology-based
ranking list and the one generated by the SIR model. The higher the
Kendall's tau value $\tau$ is, the more accurate result the method could generate. The most ideal case, $\tau = 1$,
indicates that the method uniquely identify the real influence ranking list.

\section{Numerical results}
\begin{figure}
\center{\includegraphics[scale=0.6,angle=180]{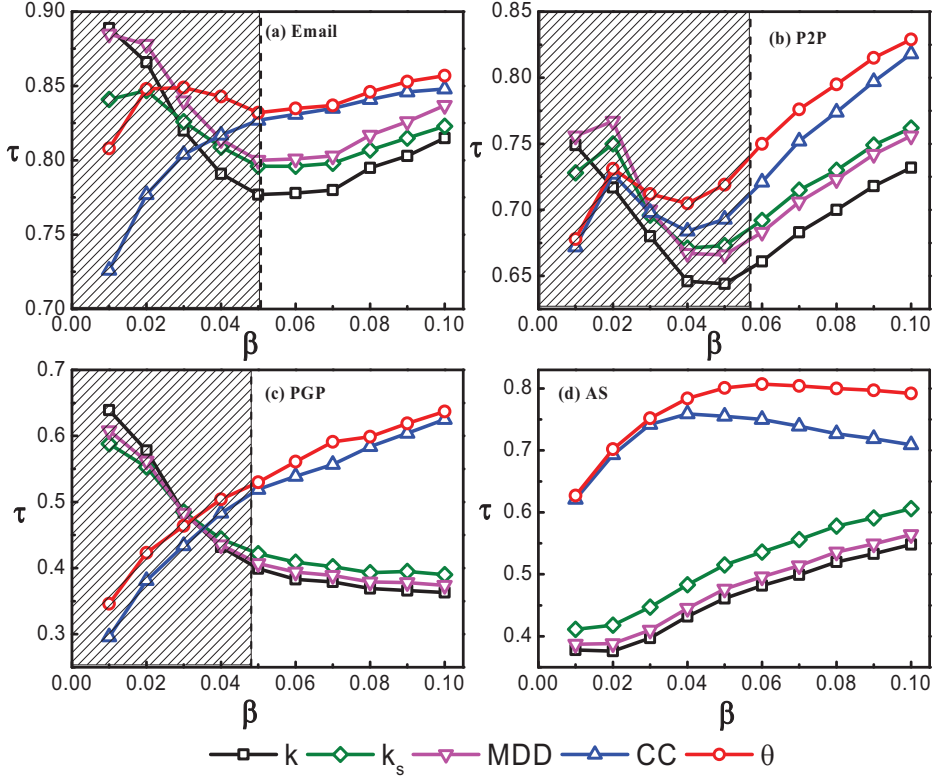}}
\caption{(Color online) The Kendall's tau values $\tau$ obtained by comparing the ranking list generated by the SIR spreading process and
the ranking lists generated by the improved method $\theta$ for Email, P2P, PGP,
and AS networks, where the dot line corresponds to the threshold $\beta^c_{\rm rand}$. From which one can find that the new method
could identify the node spreading influence more accurately when the
spreading rate $\beta$ is larger than the threshold $\beta^c_{\rm rand}$. The results are averaged over
100 independent runs with different spreading rate $\beta$.} \label{Fig2}
\end{figure}

For Email, P2P, PGP, and AS networks, the Kendall's tau values $\tau$ for the degree $k$, CC, $k$-core and
MDD indices are shown in Fig. \ref{Fig2}, from which one
can find that, when the spreading rates $\beta$ is higher than the epidemic threshold $\beta^c_{\rm rand}$ (the dot line),
the Kendall's tau value $\tau$ of the improved method $\theta$ would be much better than the other
indices.
When the spreading rate $\beta$ is much smaller than the epidemic threshold $\beta_{\rm rand}$,
the SIR process would stop in a few first infection steps, therefore the node with large degree $k$ would infect more nodes,
which may be the reason why the $\tau$ of the degree $k$ is very large when the spreading rate $\beta$ is much
smaller than $\beta^c_{\rm rand}$. The comparisons between the SIR model and the improved method show that the nodes
who are closer to the network core have more large
spreading influences, which is consistent with the main idea.

\begin{figure}
\center{\includegraphics[scale=0.6,angle=180]{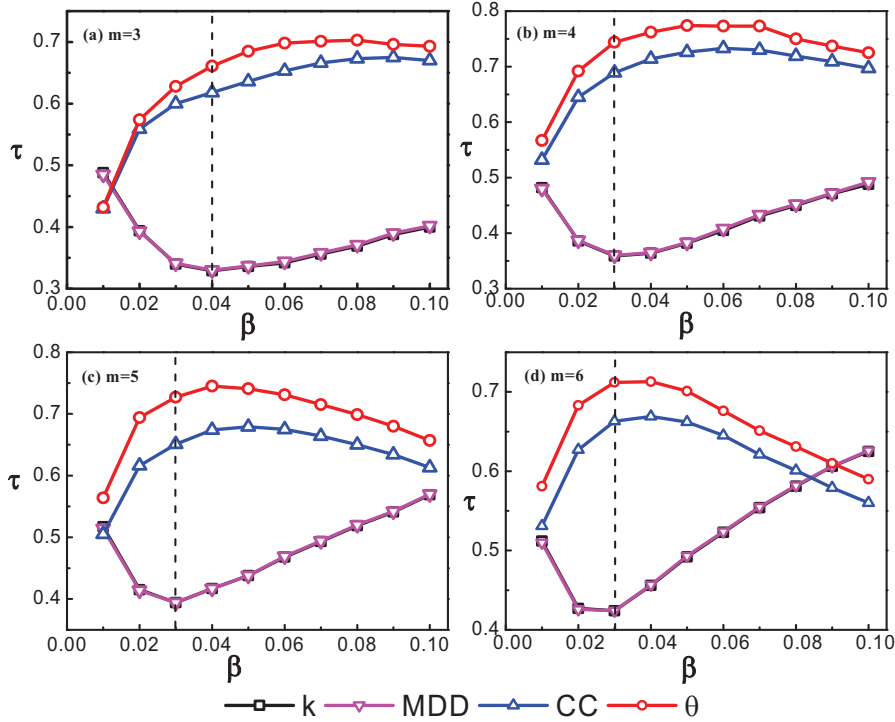}}
\caption{(Color online) Comparing the influential ranking lists generated by the improved method and the SIR result,
the Kendall's tau values $\tau$ for different BA
networks, where the dot line corresponds to the threshold $\beta^c_{\rm rand}$. The parameters
are set as $N$=7000, $m_{0}=20$ and $k_s=m=3, 4, 5, 6, 7$. From which one can find that the $\tau$ values of the degree and MDD methods are equal, which indicates that the MDD method degenerates to the degree $k$ for the BA networks, and the $\tau$ values of the improved method are much larger than the degree $k$ and MDD methods when the spreading rate $\beta$ is larger than the threshold $\beta^c_{\rm rand}$ and smaller than the crossover. The results are averaged over 100 independent runs with different
spreading rates.} \label{Fig4}
\end{figure}

Besides the real networks, we also investigate the performance of the
improved method $\theta$ for the Bab\'{a}si-Albert network \cite{14,BA}, namely BA
network. In the BA network, there are $m_0$ nodes in initial condition. In each time step, a new
node with $m$ links would connect the existed nodes according to the preferential attachment mechanism.
By using the $k$-shell decomposition method for a BA network, one can find that
all nodes have the same $k_s=m$ values except the initial $m_0$ nodes, which indicates that the traditional $k$-shell method
could not be used to analyze this kind of networks. By implementing our improved method, one can find
that the node spreading influence could be ranked more accurately than the degree $k$ and MDD methods.
Figure \ref{Fig4} shows
that the Kendall's tau values $\tau$ generated by the degree $k$ and MDD methods are equal for different spreading rate $\beta$ for the BA networks, which indicates that the MDD method would degenerate to the degree $k$ index for the BA network. One also could find that,
when the spreading rate $\beta$ is larger than the spreading threshold $\beta^c_{\rm rand}$, the Kendall's tau values $\tau$ would increase accordingly. It should be emphasized that when $\beta$ is very large, there would be a crossover
for the Kendall's tau values generated by the $\theta$ and the ones generated by the degree $k$, MDD methods.
The reason may be lie in the fact when the spreading rate $\beta$ is very large, the network core would be very easy to be infected by the nodes with largest degrees, and then the infections would be spread to the entire network very quickly.

In order to investigate the capability of the method to distinguish the spreading influences of the nodes with same $k$-core values,
we define the {\it distinct} metric $D$ as follows
\begin{eqnarray}\label{eqn2}
D=\sum_{k_s^{\rm min}}^{k_s^{\rm max}}\frac{\#{\rm distinct\ elements \ in}\ S_{k_s}}{N} .
\end{eqnarray}
where $k_s^{\rm min}$ and $k_s^{\rm max}$ denote the minimum and maximum $k$-core values
of a network. For the nodes belongs to the node set $S_{k_s}$, the number of distinct elements could be used to measure the improvement of
the method to the traditional $k$-shell decomposition method. For example, for a node set $\{1,2,3\}\in S_1$ whose degrees are 5,2,5, one can find
that there are two distinct degree values $\{2,5\}$, therefore, the number of distinct elements for $S_1$ equals to two. The largest value $D=1$
indicates that all nodes of each node set $S_{k_s}$ ($k_s \in [k_s^{\rm min}, k_s^{\rm max}]$) could be identically distinguished, while the minimum value $D=\frac{1}{N}$ means that all nodes are assigned the same $k_s$ value.

Table \uppercase\expandafter{\romannumeral2} shows the results of
the distinct $D$ for four real networks, which indicate
that, comparing with the degree $k$ and MDD methods, our improved method
$\theta$ could have much better ability of distinguishing the node spreading influences for the
nodes with same $k_s$ value.

\begin{table}
\caption{The distinct $D$ values for Email, P2P, PGP and AS networks.}
\begin{center}
\begin{tabular} {ccccc}
  \hline \hline
   Index      & Email        & P2P          & PGP          & AS   \\ \hline
   $k$         &  12.27\%      & 2.57\%       &3.76\%        &3.58\%   \\
   MDD         &  20.92\%      & 5.33\%       &3.57\%        &2.95\%   \\
   $\theta$    &  {\bf 42.81\%} & {\bf 28.92\%} &{\bf 13.55\%}  &{\bf 4.60\%}    \\
   \hline \hline
   \end{tabular}
\end{center}
\end{table}

\section{Conclusions an discussions}
In summary, we propose a parameterless method to rank the node spreading influence in terms of the
node distance to the network core which is defined as the node set with highest $k_s$ values.
The $k$-shell decomposition method could identify the most
influential spreaders of a network, and also assign some nodes with the
the same value regardless their characters in the spreading process. According to the SIR spreading process results,
one can find that the nodes with the same $k_s$ value
have far different spreading influences. The nodes whose locations are close
to the network core play more significant role in the spreading process. Taking into
account the $k_s$ values and the shortest distance between the target node and the network core, we
propose an improved method to rank the node spreading influence. The simulation
results for four real networks and the BA network show that, comparing
with the SIR spreading process results, this method could identify the
node spreading influence more accurately than the degree $k$, closeness centrality, $k$-shell and MDD decomposition methods.
Our method is parameterless and only
depends on the $k_s$ value and the distance from the target node to the network core, which is very helpful for
the widely application in real systems.

Being great value to practice and theory, several methods are
proposed to rank the node spreading influence. However, presented methods
so far mainly focus on the node degrees or positions.
We here turn to a new perspective to understand the relationship
between not only the $k$-shell location, but the nodes' shortest
distance to the network core.
Up to now, although the distance-based
method could shed some light on how the position and
distance to the network core affect the node spreading influence, we
still lack systematic comparison and understanding of the
performances of these measures, which is set as our future work.
Klemm {\it et al.} \cite{r01} argued that the importance of a node in a network is not uniquely determined by the system structure, but
it is a result of the interplay between dynamics and network structure.
Empirical analysis on more known and proposed indices as well
as more dynamic models \cite{PRL,Liu2006} is very valuable for
deeply understanding the spreading dynamic and building up knowledge
and experience. A clear picture of this issue can be completed by
putting together of many fragments from respective empirical
studies. Besides the empirical results, an alternative way is to
build artificial network models with controllable topological
features. In this way, we could have a clear picture on the unknown
and uncontrollable ingredients which are always mixed together in
real networks.

\section*{acknowledgments}
This work is partially supported by NSFC (91024026, 71071098 and 71171136), the Innovation Program of Shanghai Municipal
Education Commission (11ZZ135, 11YZ110), JGL is supported by the
Shanghai Rising-Star Program (11QA1404500) and Key Project of
Chinese Ministry of Education (211057).


\begin{thebibliography}{0}
\bibitem{kendall23}
Let $(x_1, y_1), (x_2, y_2), . . . , (x_n, y_n)$ be a set of joint
observations from two random variables $X$ and $Y$ respectively,
If one has $x_i > x_j$ and $y_i > y_j$ or $x_i < x_j$ and $y_i < y_j$,
the observations $(x_i, y_i)$ and $(x_j, y_j)$ are said to be
concordant. If $x_i > x_j$ and $y_i < y_j$ or if both $x_i < x_j$
and $y_i > y_j$, the ranks for both elements are said to be
discordant. If $x_i = x_j$ and $y_i = y_j$, the pair is neither
concordant nor discordant. Then Kendall's tau could be defined as
$\frac{N_c-N_d}{\sqrt {(n_0-n_1)(n_0-n_2)}},$
where $n_0=n(n-1)/2$, $n_1=\sum_{i}t_{i}(t_{i}-1)/2$,
$n_2=\sum_{j}t_{j}(t_{j}-1)/2$, $N_c$ and $N_d$ are the number of
concordant and discordant pairs, $t_i$ and $t_j$ are the number of
tied values in the $i$th and $j$th group of ties respectively.
\end{thebibliography}
\end{document}